\documentclass[a4,11pt]{reportN}

\usepackage[utf8]{inputenc} 
\usepackage[brazilian,portuges]{babel}

\usepackage{graphics}
\usepackage{subfigure}
\usepackage{graphicx}
\usepackage{epsfig}
\usepackage[centertags]{amsmath}
\usepackage{graphicx,indentfirst,amsmath,amsfonts,amssymb,amsthm,newlfont}
\usepackage{longtable}
\usepackage{cite}
\usepackage[usenames,dvipsnames,svgnames,table]{xcolor}
\usepackage[numbers,sort&compress]{natbib}

\begin{document}
    
    %********************************************************
    \title{Criptografia de imagens baseada em computação flexível de sistemas caóticos}
    %Este trabalho foi apresentado no evento X, na cidade Y no
    %ano Z.}}
    
    \vspace{-0.5cm}
    \author
    { \Large{Rafael Cassemiro Gonzalez }\thanks{rafaelcassemirogonzalez@gmail.com} \\
        %{\small Graduação em Engenharia Elétrica, UFSJ, São João del-Rei, MG}\\
        \Large{Erivelton Geraldo Nepomuceno}\thanks{nepomuceno@ufsj.edu.br} \\
        %{\small Departamento de Engenharia Elétrica,  UFSJ, São João del-Rei, MG}\\}
        {\small Grupo de Controle e Modelagem (GCOM), UFSJ, Brasil}
    }
    
    \criartitulo
    
    %\markboth{\hfill Write the title of your work here, concisely if
    %necessary \hfill Write the last authors' name here} {Proceeding Series of the Brazilian
    %Society of Computational and Applied Mathematics \hfill}
    \vspace{-0.5cm}
    \begin{abstract}
        {\bf Resumo}. O aumento do tráfico de dados na internet ampliou significativamente a relevância de criptografia de dados e imagens. Entre as técnicas mais empregadas em criptografia, os sistemas caóticos tem recebido grande atenção devido sua fácil implementação. Entretanto, recentemente foi observado que estes sistemas podem perder suas propriedades caóticas devido a precisão finita dos computadores. Neste trabalho, pretende-se investigar a ferramentas de computação flexível, particularmente a análise intervalar, para reduzir esse problema. Optou-se por Sistema de Lorenz, pois trata-se de um dos poucos sistemas cuja caoticidade é comprovada analiticamente. Os resultados deste trabalho, pautados nos índices de correlação e entropia, foram superiores a outros três trabalhos publicados na literatura recente. 
        
        \noindent
        {\bf Palavras-chave}. Criptografia, Caos, Computação Aritmética, Limite Inferior do Erro.
    \end{abstract}
    
    \section{Introdução}
    \label{sec:Int}
    
    A segurança tecnológica se torna ainda mais relevante com a crescente necessidade de assegurar procedimentos financeiros, sendo eles empregados em softwares bancários ou a utilização de moedas digitais, tal como a \textit{bitcoin} \citep{chenaghlu2016novel}. Com isso, a capacidade de criptografar dados de forma simples e efetiva ganha o foco de muitos pesquisadores.
    
    Uma metodologia promissora para criptografia é a aplicação de sistemas caóticos na criação de chaves criptográficas, trabalhos sobre este tópico são amplamente encontrados na literatura, sejam eles atuais \cite{Ismail2018,NARDO201969}, ou da ultima década \cite{Pareek2006}. No fim da década de 1980, Henring e Palmore \cite{HP1989} afirmaram que geradores de números pseudo-aleatórios são sistemas dinâmicos caóticos \cite{Mon2002}. Tal afirmação, reforçam que sistemas como o de Lorenz \cite{Lorenz}, possuem propriedades favoráveis para gerar uma chave criptográfica. 
    
    Este trabalho pretende explorar a degradação do caos a favor da criptografia. Para isso é adotada uma análise intervalar \cite{Nep2014} com base na representação numérica normatizada pelo IEEE \cite{Ove2001,IEE2008} para o sistema de Lorenz \cite{Lorenz}. Sendo então possível concluir que órbitas caóticas, equivalentes de maneira analítica, porem com modelagens distintas, apresentam comportamento divergente quando comparadas, podendo este ser quantificado pelo dimensionamento do limite inferior do erro \cite{Nepomuceno2017}. Toma-se como hipótese que esse erro possui propriedades estatísticas adequadas para ser utilizado como gerador de números aleatórios. A ideia pode ser vista como uma técnica de \textit{soft computing} como já foi explorado em \cite{Nepomuceno2019}.
    
    O restante do trabalho está organizado da seguinte forma. Na Seção 2 é apresentada a metodologia abordada no trabalho. Em seguida, na Seção 3, os resultados obtidos são descritos e analisados. Por fim, a Seção 4 apresenta a conclusão e perspectivas de trabalhos futuros.

    %================== Conceitos Preliminares ========================
    \section{Metodologia}
    \label{sec:Cp}
    
    \subsection{O sistema de Lorenz}
    
    Em 1963 as descobertas de Edward Lorenz revelaram ao mundo o denominado atractor de Lorenz, uma serie de modelagens matemáticas que refletiam o comportamento dinâmico da atmosfera terrestre. Tal sistema é definido pelas Eqs.  \eqref{eqlo1}-\eqref{eqlo3}.
    \begin{eqnarray}
        \frac {dx}{dt} & = & \sigma (y-x) \\
        \label{eqlo1}
        \frac {dy}{dt} & = & x(\rho -z)-y\\
        \label{eqlo2}\frac {dz}{dt}& = & xy-\beta z   
        \label{eqlo3}
    \end{eqnarray}
    
    Já no final do século XX, \cite{Tuc1999a} afirma que as equações que regem o sistema Lorenz suportam um atractor estranho, com isso foi comprovado a analiticamente existência do sistema caótico de Lorenz.

    \subsection{Limite inferior do erro}
    
    O limite inferior do erro (\textit{Lower bound error}) é uma técnica que tem origem na aritmética intervalar e relaciona duas pseudo-órbitas. A partir dessa técnica é possível mensurar o erro entre a representação computacional de extensões intervalares por seus limites inferiores. O processo pode ser definido pela Eq. \eqref{eqlb}: 
    \begin{equation}
        \delta_{a,b} = \dfrac{\left|{\hat{x}_{a,n}-\hat{x}_{b,n}}\right|}{2},
        \label{eqlb}
    \end{equation}
    em que $\hat{x}_{a,n}$ e $\hat{x}_{b,n}$ são pseudo-órbitas de um sistema. Pensando no sistema de Lorenz as extensões intervalares podem ser dadas por: 
    
    \begin{equation}
        \displaystyle {{\hat{x}_{a,n}}=x(\rho -z)-y}
        \label{ext1}
    \end{equation}
    \begin{equation}
        \displaystyle {{\hat{x}_{b,n}}=x\rho -xz -y}
        \label{ext2}
    \end{equation}
    
    \subsection{Processo de Criptografia}
    
    O processo descrito em \cite{Cao2015} é composto dos seguintes passos:
    
    \begin{itemize}
        
        \item \textbf{Passo 1}: O número necessário de iterações do sistema, ou seja o número de elementos da sequencia pseudo-aleatória, pode ser encontrado através da equação: $It = {2000 + N \times M} - 1$,  em que It é o comprimento da sequencia e $M \times N$ as dimensões da imagem. 
        \item \textbf{Passo 2}: Uma vez gerada, a sequência é um vetor de dados em um determinado formato numérico, para adequação ao processo, o sinal pseudo-aleatório deve ser submetido a um processo de normalização.
        \item \textbf{Passo 3}: Por fim, deve-se utilizar o clássico algoritmo bit-XOR, tendo como entradas a chave criptográfica e a matriz de dados da imagem a ser criptografada, com isso é realizada a codificação da imagem, tornando-a não identificável.
    \end{itemize}

    \subsection{Testes de Qualidade em Criptografia}
    
    Para validar que uma imagem está criptografada de maneira satisfatória, podem ser realizados testes que evidenciem sua eficácia, sendo esses o teste que mensura o coeficiente de correlação entre píxeis adjacentes, o teste de entropia de Shannon e a análise da distribuição dos píxeis de uma imagem a partir de seu histograma. Em relação ao histograma, para uma imagem criptografada a distribuição de frequências de cores e intensidades dos píxeis deve ser uniforme, tais que não apresente qualquer destaque visual.
    
    De fato, em uma imagem genérica como uma foto, píxeis adjacentes contem alto grau de correlação entre si. No entanto, é esperado que em uma imagem criptografada a correlação entre píxeis de modo vertical, horizontal e diagonal seja próximo de zero. O coeficiente de correlação entre píxeis pode ser medido pela Equação \eqref{eqcol} \cite{Diaconu2017}. 
    \begin{equation}
        \label{eqcol}
        \rho(X,Y)=\dfrac{E[(X-\mu_X)(Y-\mu_Y)]}{\sigma_X\sigma_Y}
    \end{equation}
    \noindent em que $X$ representa a serie de píxeis da posição, $Y$ a serie de píxeis adjacentes, $\mu$ e $\sigma$ são valores de desvios real e padrão, além de $E$ ser o valor esperado matematicamente. 
    
    O teste de Shannon é uma ferramenta que mede a aleatoriedade em sistemas de comunicação. Tal teste é definido pela equação  \ref{eqent} \cite{Luo2015}:
    \begin{equation}
        \label{eqent}
        H(X)=\sum\limits_{i=1}^{2^N-1}P_ilog_2 \dfrac{1}{P_i}
    \end{equation}
    
    \noindent em que $H(X)$ é a entropia (bits), $X$ é um símbolo e $P_i$ é o valor probabilístico de $X$.

    %================== Metodologia ===================================
    \subsection{Validação}
    
    \label{sec:Met}
    O sistema de Lorenz é simulado no  \textit{Software} MATLAB, utilizando o método Runge-Kutta de quarta ordem, com passo de integração de $10^{-6}$. Seguindo os passos definidos nos na seção 2.3, para uma imagem com resolução de $256\times256$ píxeis, o numero de iterações é de $It = 67535$. A partir do limite inferior do erro entre as pseudo-órbitas é gerada a sequencia pseudo-aleatória, que será normalizada para 8 bits (base numérica da imagem a ser codificada) e em seguida utilizada como chave do processo criptográfico numa porta XOR, como indica o passo 3.
    
    Para avaliar a qualidade da criptografia serão utilizados os seguintes indicadores: coeficiente de correlação (horizonta, vertical e diagonal) e entropia. Como são valores que variam entre faixas distintas, propõem-se aqui um índice de eficiência que leva em conta o peso de quatro índices. Esse índice de eficiência é expresso pela seguinte equação:
    \begin{equation}
        \label{eqind}
        Ic_i= {Media} (\cfrac{V_{ch}^i}{M_{ch}},\cfrac{V_{cv}^i}{M_{cv}},\cfrac{V_{cd}^i}{M_{cd}},\cfrac{V_{e}^i}{M_{e}})
    \end{equation}
    
    \noindent em que $Ic_i$ é o valor de índice do trabalho a ser calculado, $V_{ch}^i$, $V_{cv}^i$, $V_{cd}^i$ e ${V_{e}^i}$ são os valores dos testes de correlação horizontal, vertical, diagonal e entropia, respectivamente de um trabalho $i$, dividido pelos respectivos melhores valores de cada teste, representados por $M_{ch}$, $M_{cv}$, $M_{cd}$ e $M_{e}$. 
    
    Como valores de $M$ próximos a zero representam bons resultados para testes de correlação, deve-se considerar o inverso do valor para o calculo do índice, já para valores de entropia, segue-se a logica normal. Os nossos resultados serão comparados com os valores obtidos pelos trabalhos \cite{NLNA2018},\cite{Li2017} e \cite{Luo2015}.
    
    %================================ Resultados ======================================
    
    \section{Resultados}
    \label{sec:Res}
    
    \begin{figure}[!h]
        \centering
        \includegraphics[width=0.7\linewidth]{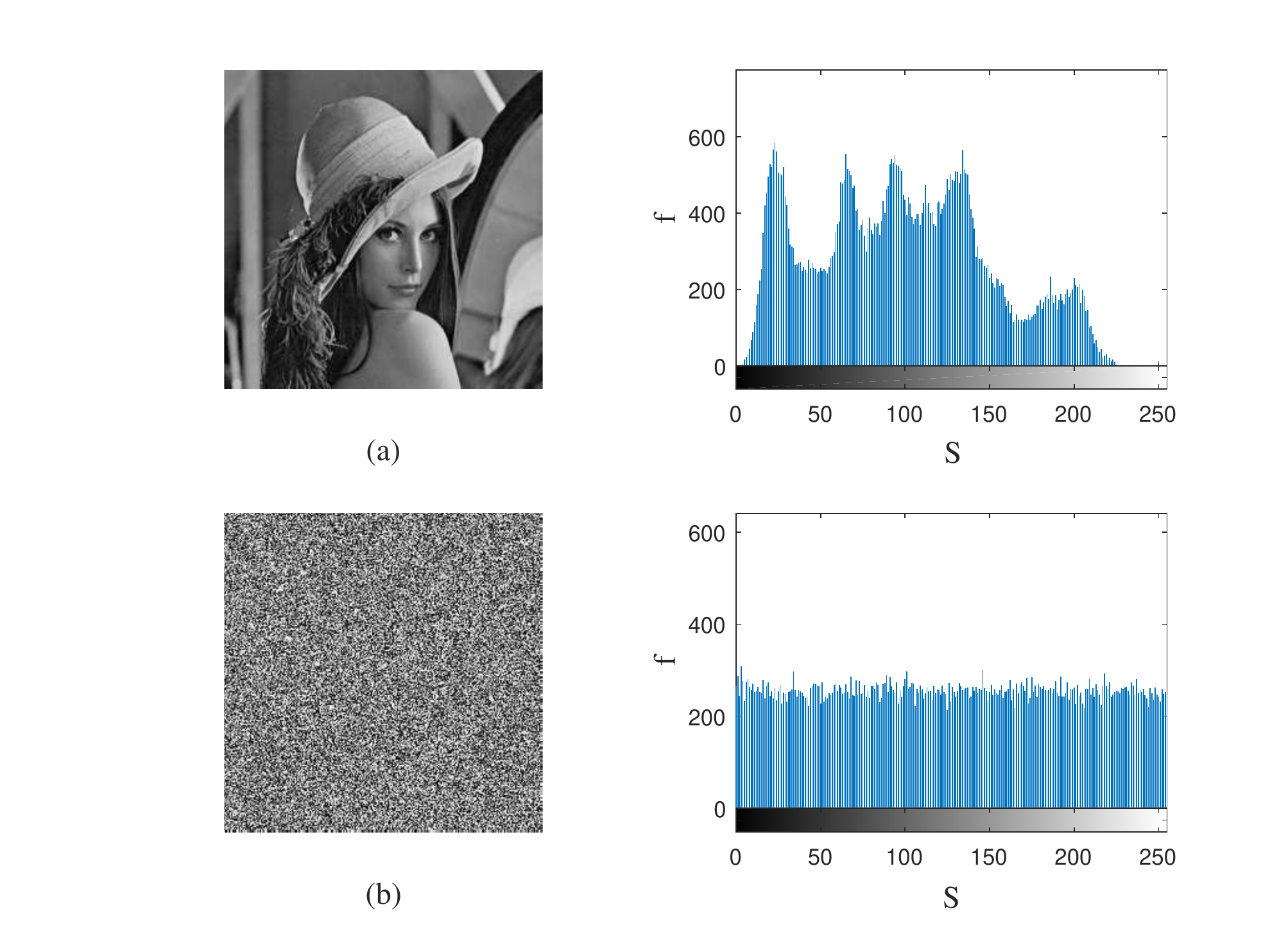}
        \caption{Comparativo de imagens e seus histogramas, sendo (a) é a imagem original e (b) é a imagem criptografada. Os respectivos histogramas ao lado direito de cada imagem, apresentam nos seus eixo f a frequência de repetibilidade de uma tom de cinza e o eixo S representa a escala de cinza das figuras. É visível a eficácia do processo ao comparar a imagem criptografada à original, além da distribuição homogênea de seu histograma após o processo, validando a metodologia.}
        \label{fig:TOT}
    \end{figure}

    A imagem base do processo está presente figura \ref{fig:TOT} (a), Lena, esta é comumente utilizada no campo de processamento de imagens devido ao seu detalhamento, sombreado e textura. Ao montar o histograma (presente na figura \ref{fig:TOT} a direita de Lena) da imagem é possível observar a frequência de determinada tonalidade da escala de cinza, o que mostra um espectro bem distribuído e heterogêneo de tons de cinza.

    % \begin{figure}[!h]
    %     \centering
    %     \includegraphics[width=0.5 \linewidth]{LB.eps}
    %     \caption{Erro de limite inferior na base $log_{10}$, em que $n$ representa o número de interações. O crescimento do erro pode ser analisado como rampa até pouco depois de 2000 iterações, acima deste valor o erro oscila de forma aproximada a uma reta constante.}
    %     \label{fig:LB}
    % \end{figure}
    
    Para a operação do sistema foram escolhidos arbitrariamente os valores das constantes $\sigma = 16$ , $\rho=45,92$ e $\beta = 4$. Além disso, foram estabelecidas as constantes iniciais de cada órbita, sendo $V_1 = 1$, $V_2 = 0,5$ e $V_1 = 0,9$. Tais entradas resultam em pseudo-órbitas que aplicadas á Equação \eqref{eqlb} originam o limite inferior do erro. Após a utilização do sinal como chave criptográfica, a imagem original foi convertida na imagem \ref{fig:TOT} (b), nesta não é possível identificar qualquer traço da imagem original e ainda apresenta uma distribuição relativamente homogênea de tons de cinza, como pode ser observado pelo seu histograma (à direta).
    
    \begin{table*}[]
        \caption{Coeficiente de correlação (Vertical, Horizontal e Diagonal, respectivamente nas três primeiras colunas) e entropia em comparativo com outros trabalhos. Na tabela é possível observar que os valores de entropia e os coeficientes de correlação estão em uma margem satisfatória em relação a outros trabalhos já publicados. O índice de eficiência retrata a média da razão de cada resultado em relação ao melhor valor em sua categoria, conforme estabelecido na Eq. \eqref{eqind}, a análise comparativa dos índices indica que a proposta apresenta valor superior as outros três trabalhos de referência.}
        \begin{center}
            \begin{tabular}{cccccc}
                \hline
                %\multicolumn{3}{|c|}{Coeficiente de Correlação} & %\multirow{}{}{Entropia} & \multirow{}{}{Índice de %eficiência} & \multirow{}{}{Referências}} \\ \cline{1-3}
                %Horizontal} & Vertical} & Diagonal} &  &  &  \\ \hline
                Hor. & Ver. & Dia. & Entropia & Eficiência & Ref. \\ \hline
                0,00045 & 0,0015 & 0,0040 & 7,9973 & 0,7686 & Este trabalho \\ 
                0,0028 & 0,0059 & 0,0031 & 7,9969 & 0,3778 & Nardo et. al \cite{NLNA2018} \\ 
                0,00083 & 0,00223 & 0,00650 & 7,9998 & 0,5652 & \cite{Li2017} \\
                0,0016 & 0,0025 & 0,0003 & 7,9826 & 0,7197 & \cite{Luo2015} \\ \hline
            \end{tabular}
        \end{center}
        \label{tab1}
    \end{table*}
    
    Em uma nova aplicação da mesma chave criptográfica, observa-se o processo de decriptografia, no qual a imagem criptografada retorna para sua forma original, mantendo sua distribuição de escalas de cinza e todas suas características visuais, sendo assim comprovada a funcionalidade do processo criptográfico geral. Para fins comparativos entre trabalhos semelhantes \cite{NLNA2018,Li2017,Luo2015},  a tabela \ref{tab1}  apresenta dados de testes de coeficiente de correlação e entropia e o índice de eficiência referenciado pela Equação \eqref{eqind}. A análise dos resultados evidencia que o processo tem valores satisfatórios, em relação ao outros trabalhos, além disso, comparando os valores do índice, é possível verificar uma superioridade relativa deste trabalho.

    \section{Conclusão}
    \label{sec:Con}
    
    Neste trabalho foi proposto um novo método de criptografia que utiliza o erro de pseudo-órbitas do sistema de Lorenz na criação de uma chave com propriedades estatísticas adequadas para realizar a criptografia de imagens. Os dados apresentados na tabela \ref{tab1} evidenciam que a eficiência está no mesmo nível de outros trabalhos presentes na literatura.
    
    A perda de propriedades caóticas foi obtida a partir de técnicas de computação flexível, nas quais determinam o limite do inferior do erro que pode ser visto como uma estimativa do intervalo inferior da propagação do erro. Como aspiração para trabalhos futuros é proposto a realização de outros testes, tal como o NIST SP 800-22 \cite{Bassham2010,Tutueva2017}  para o aprofundamento em análise do desempenho do método.   
    
    \section*{Agradecimentos}
    Agradecemos a CAPES, CNPq e Fapemig pelo suporte financeiro. Ao GCOM pelo apoio.

    %\bibliographystyle{unsrtnat}
    %\bibliography{19sbai-rafael}

\end{document}